\documentclass{PoS}

\newcommand{\bc}{\begin{center}}
\newcommand{\ec}{\end{center}}
\newcommand{\beqn}{\begin{equation}}
\newcommand{\eeqn}{\end{equation}}
\newcommand{\barr}{\begin{eqnarray}}
\newcommand{\earr}{\end{eqnarray}}

\def\simge{\mathrel{%
       \rlap{\raise 0.511ex \hbox{$>$}}{\lower 0.511ex \hbox{$\sim$}}}}
\def\simle{\mathrel{
       \rlap{\raise 0.511ex \hbox{$<$}}{\lower 0.511ex \hbox{$\sim$}}}}

\def\SR    {\mathord{\buildrel{\lower3pt\hbox{$\scriptscriptstyle\rightarrow$}}\over S}}
\def\SL    {\mathord{\buildrel{\lower3pt\hbox{$\scriptscriptstyle\leftarrow$}}\over S}}

\def\simge{\mathrel{%
       \rlap{\raise 0.511ex \hbox{$>$}}{\lower 0.511ex \hbox{$\sim$}}}}
\def\simle{\mathrel{
       \rlap{\raise 0.511ex \hbox{$<$}}{\lower 0.511ex \hbox{$\sim$}}}}
\title{Thermodynamics and heavy-quark free energies
 at finite temperature and density 
 with two flavors of improved Wilson quarks}

\ShortTitle{Thermodynamics and heavy-quark free energies 
with improved Wilson quarks}

\author{\speaker{Y.~Maezawa}, T.~Hatsuda\\
        Department of Physics, The University of Tokyo,\\
        Bunkyo-ku, Tokyo 113-0033, Japan\\
        E-mail: \email{maezawa@nt.phys.s.u-tokyo.ac.jp}}

\author{S.~Aoki, K.~Kanaya\\
        Graduate School of Pure and Applied Sciences, University of Tsukuba,\\
        Tsukuba, Ibaraki 305-8571, Japan}

\author{S.~Ejiri\\
        Physics Department, Brookhaven National Laboratory,\\
        Upton, New York 11973, USA}

\author{N.~Ishii, N.~Ukita and T.~Umeda\\
        Center for Computational Sciences, University of Tsukuba,\\
        Tsukuba, Ibaraki 305-8577, Japan}

\abstract{
Thermodynamics of two-flavor QCD at finite temperature and density 
is studied on a $16^3 \times 4$ lattice, 
using a renormalization group improved gauge action 
and the clover improved Wilson quark action. 
In the simulations along lines of constant $m_{\rm PS} / m_{\rm V}$,
 we calculate the Taylor expansion coefficients of 
 the heavy-quark free energy 
 with respect to the quark chemical potential 
 ($\mu_q$) up to the second order.
 By comparing the expansion coefficients of the free energies
 between quark($Q$) and antiquark($\bar{Q}$), 
 and between $Q$ and $Q$,
 we find a characteristic difference at finite $\mu_q$ 
  due to the first order coefficient of the Taylor expansion.
 We also calculate the quark number and isospin susceptibilities,
 and find that 
 the second order coefficient of the quark number susceptibility 
 shows enhancement around the pseudo-critical temperature.
}

\FullConference{The XXV International Symposium on Lattice Field Theory\\
		July 30 - August 4 2007\\
		Regensburg, Germany}

\begin{document}

%%%%%%%%%%%%%%%%%%%%%%%%%%%%%%%%%%%%%%%%%%%%%%%%%%%%%%%%%%%%%%%%%%%%%
\section{Introduction}

We study QCD thermodynamics 
 at finite temperature $(T)$ and quark chemical potential ($\mu_q$)
with two-flavors of dynamical quarks.
Then, we calculate the Taylor expansion coefficients of 
physical quantities in the simulations at $\mu_q=0$,
and investigate thermodynamic properties at small $\mu_q$ region.
In this report,
we present current status of two topics: heavy-quark free energy and 
fluctuation at finite $\mu_q$.
The former is related to the inter-quark interaction in 
quark-gluon plasma,
 and the latter is related to 
 the existence of the critical point in $(T, \mu_q)$ plane.

%%%%%%%%%%%%%%%%%%%%%%%%%%%%%%%%%%%%%%%%%%%%%%%%%%%%%%%%%%%%%%%%%%%%%
\section{Lattice action and simulation parameters}

We employ a renormalization group improved gauge action and a clover improved
Wilson quark action with two flavors.
The numerical simulation is performed on a lattice with a size of 
$N_s^3 \times N_t = 16^3 \times 4$
 along lines of constant physics, 
i.e. lines of constant $m_{\rm PS} / m_{\rm V}$
(the ratio of pseudoscalar and vector meson masses) at $T=0$
in the space of simulation parameters.
Two values of $m_{\rm PS} / m_{\rm V}$ are taken: 0.65 and 0.80
with the temperature range of $T/T_{pc} \sim$ 0.82--4.0
and 0.76--3.0, respectively, where $T_{pc}$ is the pseudo-critical 
temperature along the line of constant physics.
The number of trajectories for each run after thermalization is 
5000--6000, and we measure physical quantities at every 10 trajectories.
Details of the lines of constant physics with the same actions are
summarized in Ref.~\cite{cp,whot}.
To calculate derivatives of the quark determinant
with respect to $\mu_q$,
 we use the random noise method
introduced in Ref.~\cite{ejiri} with the number of noise of 100--200.

%%%%%%%%%%%%%%%%%%%%%%%%%%%%%%%%%%%%%%%%%%%%%%%%%%%%%%%%%%%%%%%%%%%%%
\section{Heavy quark free energy and Debye screening mass}

The heavy-quark free energy in the QCD medium is one of 
the important quantities to characterize the properties of 
the quark-gluon plasma.
Especially, the properties of the free energies at finite 
$T$ and $\mu_q$
may be related to the fate of the charmoniums and bottomoniums
in relativistic heavy ion collisions.
Precise studies of the heavy-quark free energy in two-flavors QCD
have been done with the improved Wilson quark action
 at $\mu_q =0$ \cite{whot}.
Properties of the free energy
 between static-quark ($Q$) and -antiquark $(\bar{Q})$
 at finite $\mu_q$
have been previously investigated by using the Taylor expansion method
using an improved staggered quark action \cite{ks1}.
In this section, we show the Taylor expansion coefficients
of the free energy 
between not only $Q$ and $\bar{Q}$, but also $Q$ and $Q$,
up to 2nd-order of $\mu_q$.

The free energy of static quarks on the lattice
is described by the correlations of the  Polyakov loop:
$\Omega ( {\bf x} )  = \prod_{ \tau = 1}^{N_t} U_4 (\tau, {\bf x})$
 where the $U_\mu (\tau, {\bf x}) \in$ SU(3) is the link variable.
 With an appropriate gauge fixing (e.g. the Coulomb gauge fixing),
one can define the free energy in various 
 color channels separately:
 the color singlet $Q\bar{Q}$ channel ({\bf 1}),
 the color octet $Q\bar{Q}$ channel ({\bf 8}),
 the color anti-triplet $QQ$ channel (${\bf 3^*}$),
 and the color sextet $QQ$ channel ({\bf 6}),
 given as follows.
\barr
e^{-F^{\bf 1}(r,T)/T}
 &=&
  \frac{1}{3} \langle {\rm Tr} 
\Omega^\dagger({\bf x}) \Omega ({\bf y})
\rangle
, 
\label{eq:f1}
\\
e^{-F^{\bf 8}(r,T)/T}
 &=& 
\frac{1}{8} \langle {\rm Tr} \Omega^\dagger({\bf x})
{\rm Tr} \Omega ({\bf y})
\rangle
-
\frac{1}{24} \langle {\rm Tr} \Omega^\dagger({\bf x})
 \Omega ({\bf y}) \rangle
\label{eq:f8}
, \\
e^{-F^{\bf 6}(r,T)/T}
 &=& 
\frac{1}{12} \langle {\rm Tr} \Omega({\bf x})
{\rm Tr} \Omega ({\bf y})
\rangle
+
\frac{1}{12} \langle {\rm Tr} \Omega({\bf x})
 \Omega ({\bf y}) \rangle
\label{eq:f6}
, \\
e^{-F^{{\bf 3}^*}(r,T)/T}
 &=& 
\frac{1}{6} \langle {\rm Tr} \Omega({\bf x})
{\rm Tr} \Omega ({\bf y})
\rangle
-
\frac{1}{6} \langle {\rm Tr} \Omega({\bf x})
 \Omega ({\bf y}) \rangle
,
\label{eq:f3}
\earr
where $r = |{\bf x} - {\bf y}|$.

For $T > T_{pc}$, we introduce normalized free energies 
$(V^{\bf 1}, V^{\bf 8}, V^{\bf 6}, V^{\bf 3^*})$
such that they vanish at large distances.
This is equivalent to define the free energies by dividing
the right-hand side of Eq.~(\ref{eq:f1})--(\ref{eq:f3}) by 
$\langle {\rm Tr} \, \Omega \rangle \langle {\rm Tr} \, \Omega^\dagger \rangle$ 
for $Q \bar{Q}$ free energies
and $\langle {\rm Tr} \, \Omega \rangle^2$ for $QQ$ free energies.

The Taylor expansion of normalized free energies with respect to
$\mu_q/T$ is described as a power series up to 2nd order:
\begin{eqnarray}
V^M(r,T,\mu_q) = 
v^M_0 +
v^M_1 \left( \frac{\mu_q}{T} \right) +
v^M_2 \left( \frac{\mu_q}{T} \right)^2 +
O(\mu^3_q)
,
\label{eq:ENFE}
\end{eqnarray}
where $M$ is the color channel.
The coefficients, $v_n^M$, can be evaluated by expanding 
the quark determinant of partition function 
in powers of $\mu_q$, then the normalization of the free energies
by $\langle {\rm Tr} \Omega \rangle$ is also taken 
order by order of $\mu_q$.

One should note that
the color singlet and octet channels do not have the odd orders in
the Taylor expansion since the free energies for both channels are
symmetric with respect to $\mu_q$. 
In other words, the free energies between $Q$ and $\bar{Q}$ are
invariant under the charge conjugation.
On the other hand, the color sextet and antitriplet channels has the odd orders
since the $QQ$ free energies are not invariant under the charge conjugation.

\begin{figure}[t]
  \begin{center}
    \begin{tabular}{cc}
    \includegraphics[width=73mm]{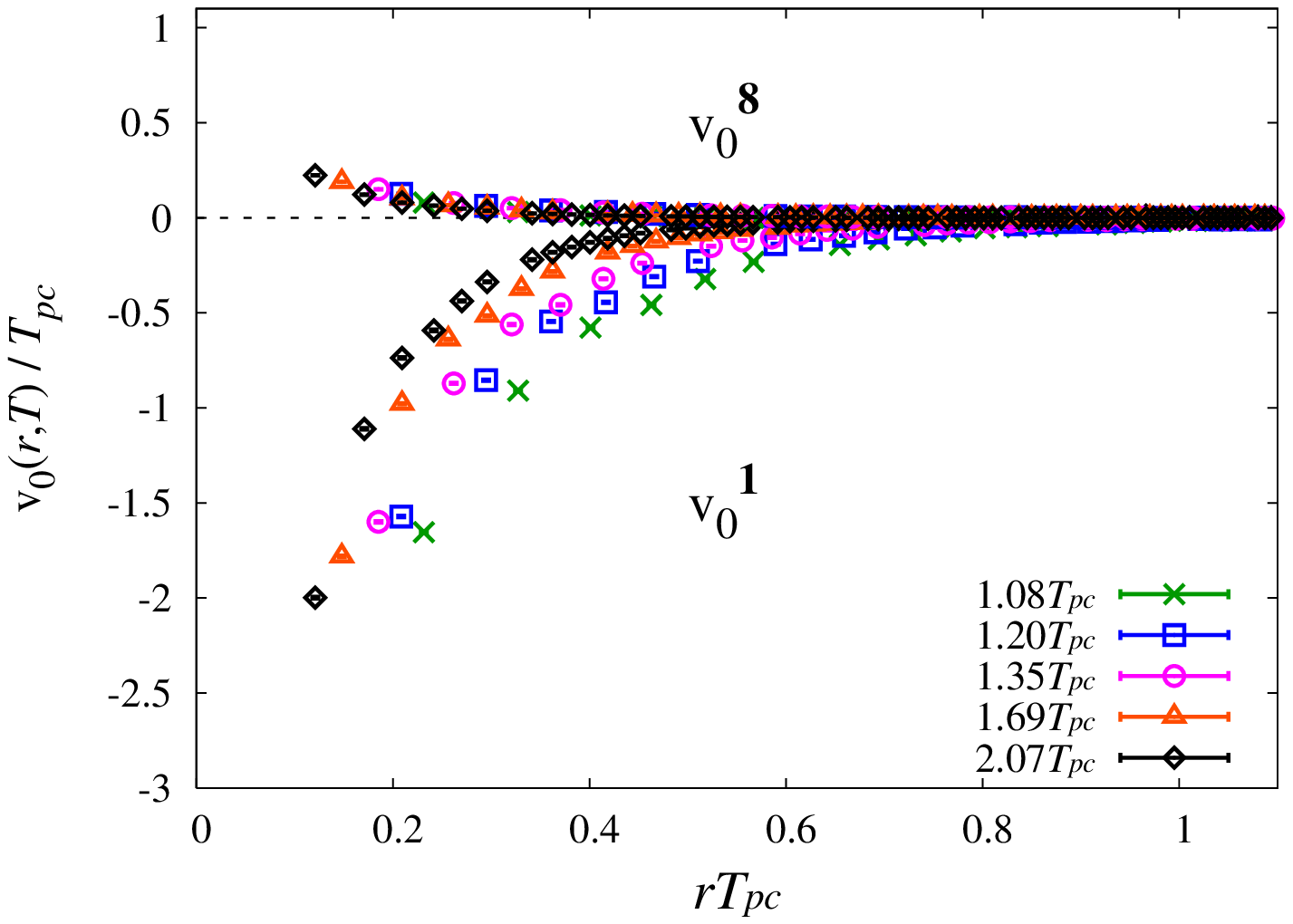} &
    \includegraphics[width=73mm]{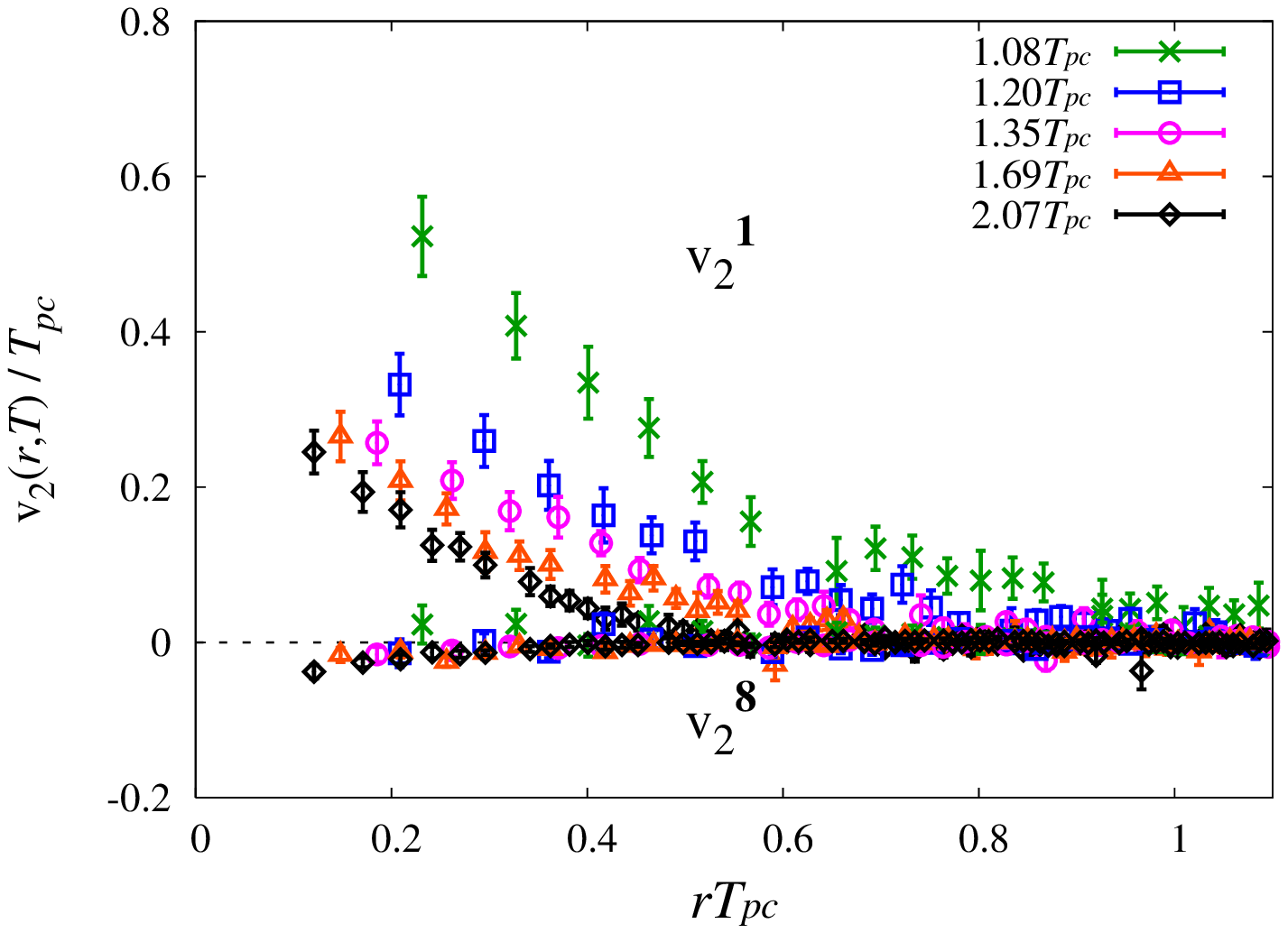}
    \end{tabular}
    \caption{Results of $v_0^M$ (left) and $v_2^M$ (right)
    for $Q \bar{Q}$ channel above $T_{pc}$
    at $m_{\rm PS}/m_{\rm V} = 0.80$.
        }
     \vspace{-0.5cm}
    \label{fig:vsi}
  \end{center}
\end{figure}

\begin{figure}[tbp]
  \begin{center}
    \begin{tabular}{cc}
    \includegraphics[width=73mm]{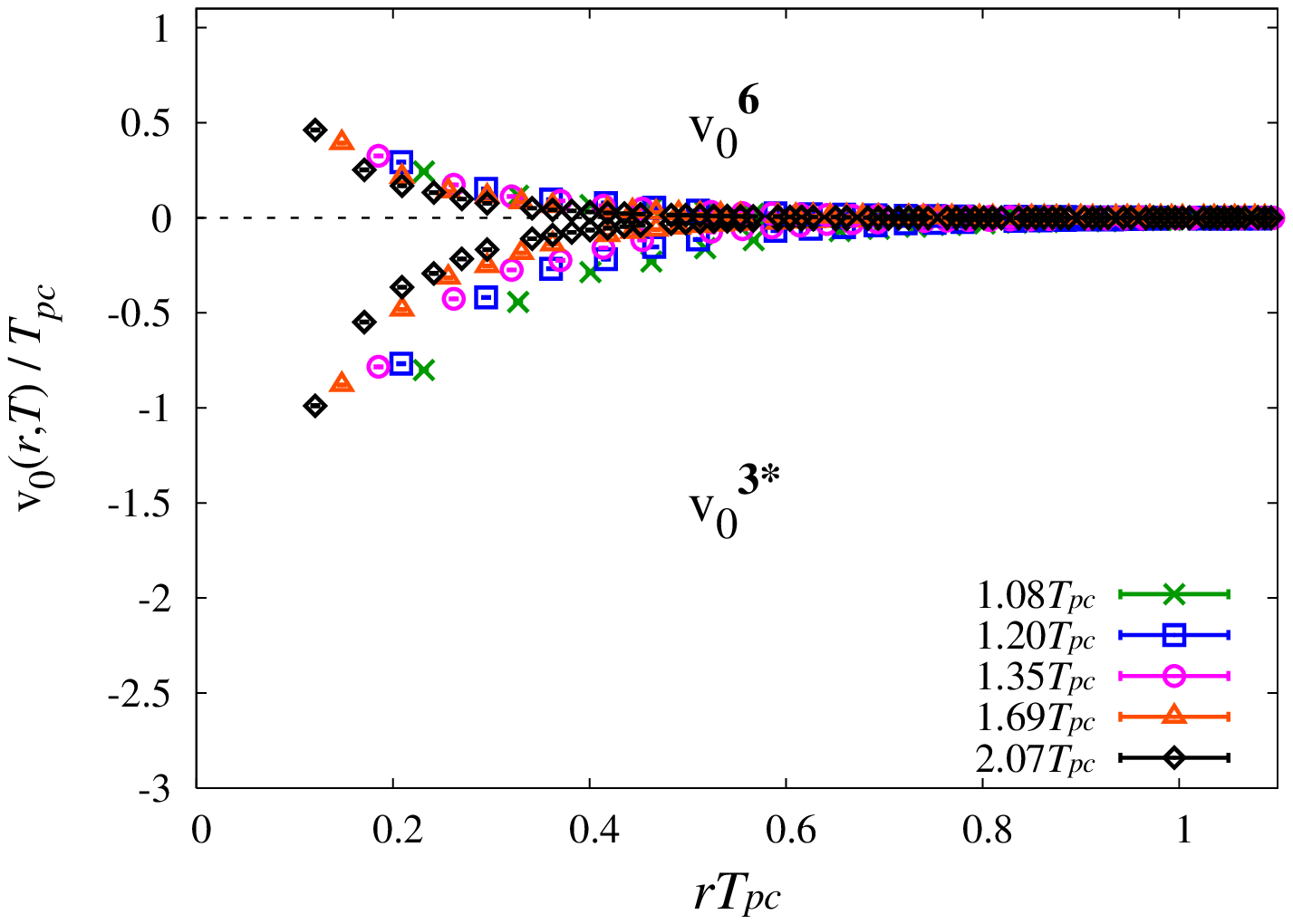} &
    \includegraphics[width=73mm]{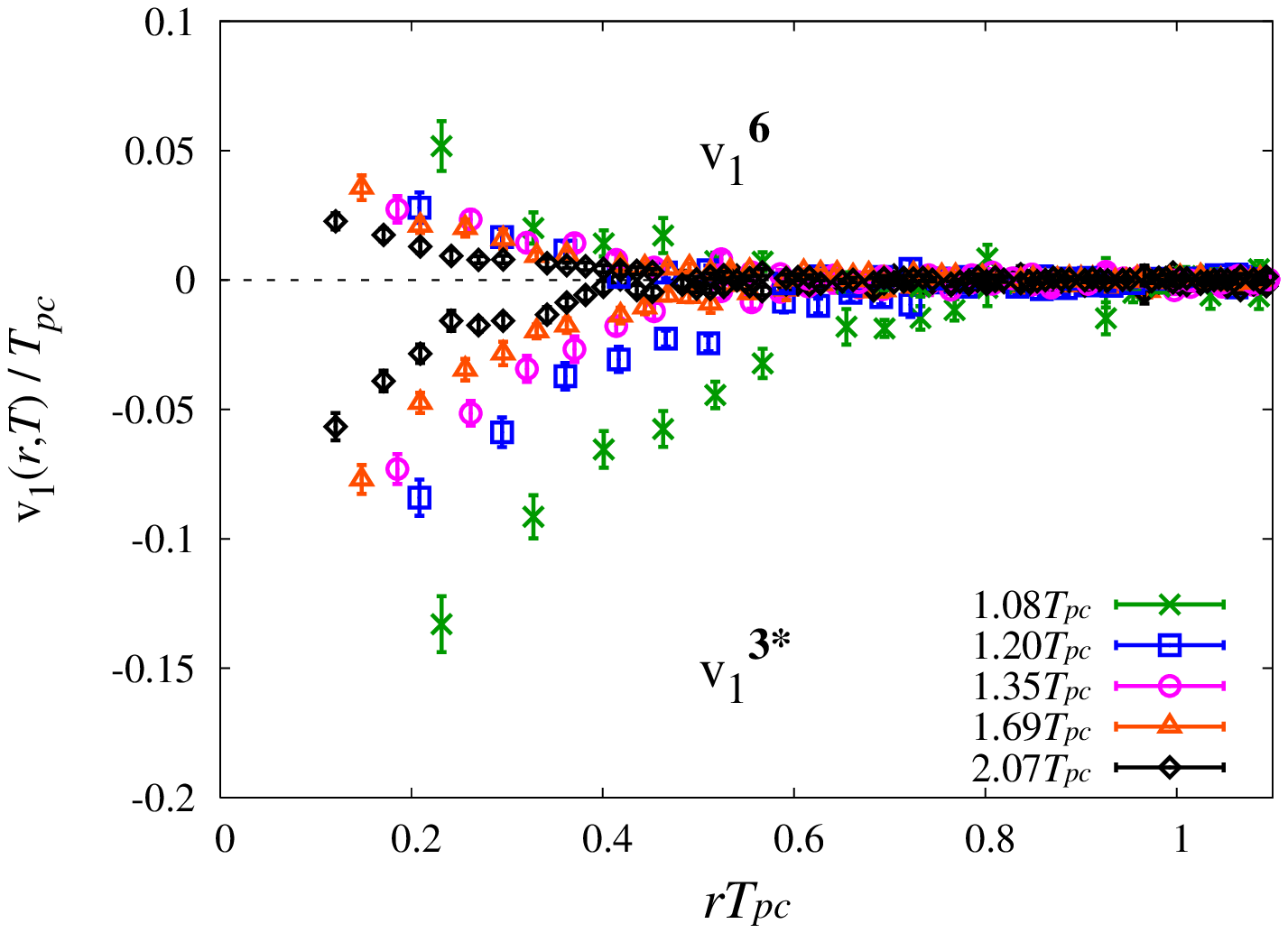}
    \end{tabular}
    \caption{Results of $v_0^M$ (left) and $v_1^M$ (right)
    for $Q Q$ channel above $T_{pc}$
    at $m_{\rm PS}/m_{\rm V} = 0.80$.
        }
     \vspace{-0.5cm}
    \label{fig:van1}
  \end{center}
\end{figure}

\begin{figure}[tbp]
  \begin{center}
    \includegraphics[width=73mm]{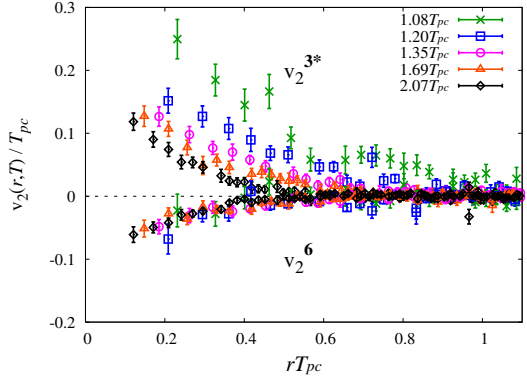}
    \caption{Results of $v_2^M$
    for $Q Q$ channel above $T_{pc}$
    at $m_{\rm PS}/m_{\rm V} = 0.80$.
        }
     \vspace{-0.5cm}
    \label{fig:van2}
  \end{center}
\end{figure}

The expansion coefficients of the normalized free energies 
for the color singlet and octet $Q \bar{Q}$ channels are 
shown in Fig.~\ref{fig:vsi} for $v_0^M$ (left) and $v_2^M$ (right)
at $m_{\rm PS} / m_{\rm V} = 0.80$ for several temperatures.
Those for the color sextet and antitriplet $QQ$ channels are
shown in Fig.~\ref{fig:van1} for $v_0^M$ (left) and $v_1^M$ (right)
and in Fig.~\ref{fig:van2} for $v_2^M$.
It have been found in Ref.~\cite{whot} that 
the inter-quark interaction is ``attractive'' in the color singlet and
antitriplet channels and is ``repulsive'' in the color octet 
and sextet channels at $\mu_q = 0$.
We find that, both in high and low temperatures, 
 the sign of $v_1^M$ is the same with that of $v_0^M$,
whereas the sign of a $v_2^M$ is the opposite of that of $v_0^M$,
i.e. 
$ v_1^M \cdot v_0^M > 0$ (only for $QQ$ free energies)
and $ v_2^M \cdot v_0^M < 0$.
This means that the inter-quark interaction between
$Q$ and $\bar{Q}$ becomes weak,
whereas that between $Q$ and $Q$ becomes strong 
in the leading-order of $\mu_q$.
In other words, $Q \bar{Q}$ ($QQ$) free energies are screened (anti-screened)
by contributions of the internal quarks induced by finite $\mu_q$.

In order to study the screening effect in each color channel,
we fit the normalized free energies by a screened Coulomb form,
\begin{eqnarray}
V^M (r,T,\mu_q) = C(M) \frac{\alpha_{\rm eff}(T,\mu_q)}{r}
  e^{- m_{D}(T,\mu_q) r}
,
\label{eq:scf}
\end{eqnarray}
where 
$\alpha_{\rm eff}(T,\mu_q)$ and $m_D(T,\mu_q)$ 
are the effective running coupling
and the Debye screening mass, respectively.
We assume that contributions of finite $\mu_q$ appear 
only in $\alpha_{\rm eff}$ and $m_D$.
The Casimir factor 
$C(M) \equiv \langle \sum_{a=1}^{8} t_1^a\cdot t_2^a \rangle_M$
 for color channel $M$  is given  by 
\begin{eqnarray}
C({\bf 1})     = -\frac{4}{3}, \quad
C({\bf 8})     =  \frac{1}{6}, \quad
C({\bf 6})     =  \frac{1}{3}, \quad
C({\bf 3^*})   = -\frac{2}{3}.
\end{eqnarray}
We assume that $m_D(T, \mu_q)$ is also expressed
as a power series of $\mu_q/T$:
\begin{eqnarray}
m_D &=& 
m_{D,0} +
m_{D,2} \left( \frac{\mu_q}{T} \right)^2 +
O(\mu^4_q)
,
\label{eq:mD}
\end{eqnarray}
where we use the fact that the Debye screening mass does not
have the odd powers in the Taylor expansion 
because it is related to the self-energy of the two-point correlation 
of the gauge fields which is symmetric when $\mu_q \rightarrow - \mu_q$.
Relations of coefficients between $V^M$ and $m_D$ are
given by comparing each order of $\mu_q$.
At $\mu_q = 0$, the relation is
 the same as that adopted in Ref.~\cite{whot}:
\barr
v_0 (r,T) &=& C(M) \frac{\alpha_{\rm eff,0} (T)}{r} e^{- m_{D,0}(T) r}
.
\label{eq:v0}
\earr
For the second order coefficients, 
we obtain,
\begin{eqnarray}
\frac{v_2}{v_0} &\simeq& - m_{D,2} \, r
.
\label{eq:v2}
\end{eqnarray}
At large distances, we estimate 
the coefficients of Debye mass by fitting the normalized
free energies for each color channel with the formulae 
(\ref{eq:v0}) and (\ref{eq:v2}).

\begin{figure}[tbp]
  \begin{center}
    \begin{tabular}{cc}
    \includegraphics[width=73mm]{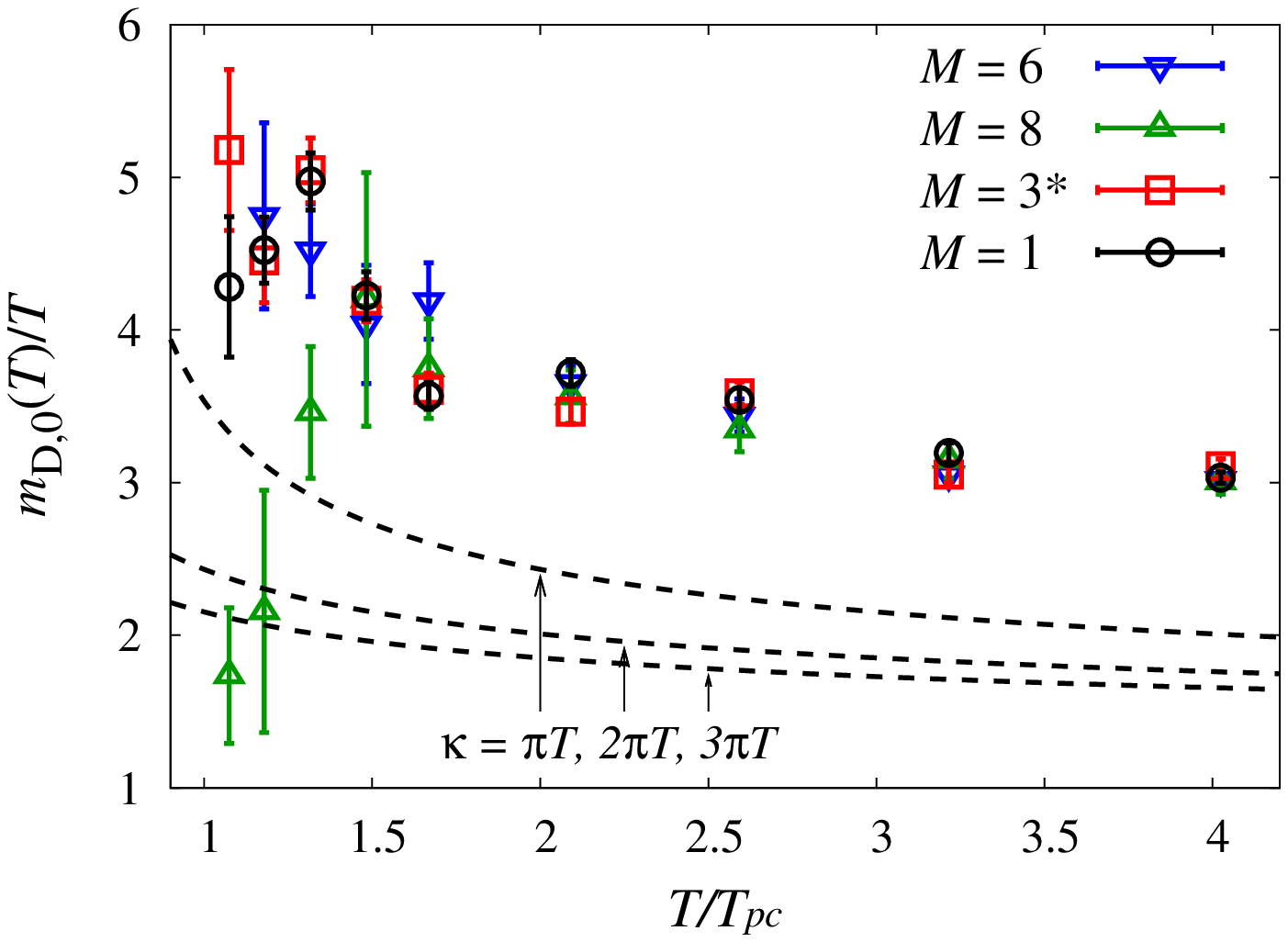} &
    \includegraphics[width=73mm]{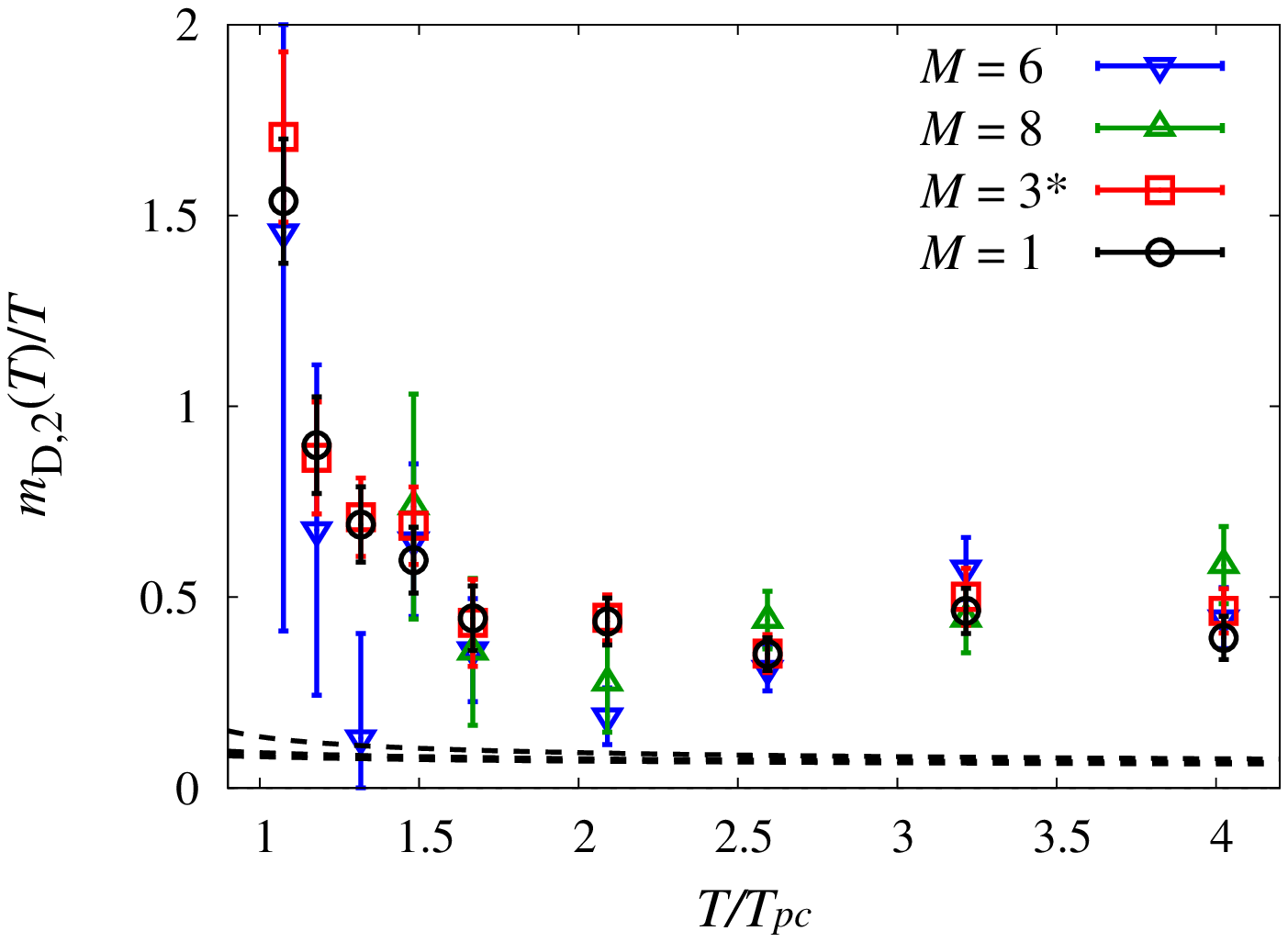}
    \end{tabular}
    \caption{Results of $m_{D,0}$ (left) and $m_{D,2}$ for each color
    channel
    at $m_{\rm PS}/m_{\rm V} = 0.65$.
    Dashed lines are prediction from a leading-order of the thermal 
    perturbation theory with the renormalization points of 
    $\kappa = \pi T$, $2 \pi T$ and $3 \pi T$.
        }
    \label{fig:mD}
  \end{center}
     \vspace{-0.5cm}
\end{figure}

 Figure \ref{fig:mD} 
 shows the results of 
the $m_{D,0}(T)$ (left) and $m_{D,2}(T)$ (right)
 for $m_{\rm PS}/m_{\rm V} = 0.65$.
Similar behavior in both results are obtained 
for $m_{\rm PS}/m_{\rm V} = 0.80$.
We find that 
there is no significant channel dependence in both coefficients
at sufficiently high temperatures $(T \simge 2 T_{pc})$.
In other words, the channel dependence of the free energy
at high temperature can be well absorbed in 
the Casimir factor.

Let us compare the Debye screening mass at finite $\mu_q$ on the lattice 
with that predicted in the thermal perturbation theory.
The 2-loop running coupling is given by
\begin{eqnarray}
g^{-2}_{2 {\rm l}} (\kappa) 
=  \beta_0 \ln \left( \frac{\kappa}{\Lambda} \right)^2 + 
\frac{\beta_1}{\beta_0} 
\ln \ln \left( \frac{\kappa}{\Lambda} \right)^2
,
\label{eq:RC}
\end{eqnarray}
where $\kappa$ is the renormalization point.
The argument in the logarithms can be written as 
$ \kappa / \Lambda = (\kappa/T)(T/T_{pc})(T_{pc}/\Lambda)$
with
$\Lambda= \Lambda_{\overline{MS}}^{N_f=2} \simeq 261$ MeV
 \cite{Gockeler:2005rv} 
and $T_{pc} \simeq 171$ MeV \cite{cp}.
We assume $\kappa$ to be in a range $\pi T$
 to $3 \pi T$.
Therefore, $g_{\rm 2l}$ can be viewed as a function of $T/T_{pc}$. 
In the leading order of the thermal perturbation theory,
the Debye screening mass with $g_{\rm 2l}$ is given by 
\begin{eqnarray}
m_D^{\rm LO} (T,\mu_q) = g_{2 {\rm l}} (T) \left\{ ( 1+ \frac{N_f}{6} )T^2
+ \frac{N_f}{2 \pi^2} \mu_q^2 \right\}^{1/2}
.
\end{eqnarray}
Therefore, the leading-order expansion coefficients in the thermal
perturbation theory are given by
\begin{eqnarray}
m_{D,0}^{\rm LO} (T) = \sqrt{1 + \frac{N_f}{6}} g_{2 {\rm l}} (T) \, T
, \ \ \ \
m_{D,2}^{\rm LO} (T) = \frac{1}{4 \pi^2} \frac{N_f}{\sqrt{1 + N_f/6}} 
g_{2 {\rm l}} (T) \, T
.
\end{eqnarray}

The dashed lines in Fig.~\ref{fig:mD} are 
the results of $m_{D,0}^{\rm LO}$
and $m_{D,2}^{\rm LO}$
for $\kappa = \pi T$, $2\pi T$ and $3 \pi T$.
We find 
that the lattice results are larger than the leading-order
thermal perturbation 
for both coefficients of the Debye mass.
Since it is known that the next-to-leading-order 
in the thermal perturbation
 with respect to $T$
 compensates for such discrepancy in the case 
at $\mu_q = 0$ \cite{whot}, 
the higher order contributions
in the perturbation theory with respect to $\mu_q$ 
could help us to understand our results on the lattice.

%%%%%%%%%%%%%%%%%%%%%%%%%%%%%%%%%%%%%%%%%%%%%%%%%%%%%%%%%%%%%%%%%%%%%
\section{Hadronic fluctuations at finite $\mu_q$}

 Hadronic fluctuations at finite density are observables closely related to 
the critical point in the $(T, \mu_q)$ plane and may be experimentally 
detected by an event-by-event analysis of heavy ion collisions. 
 The fluctuations can also be studied by numerical simulations of 
lattice QCD calculating the quark number and isospin susceptibilities,  
$\chi_q$ and $\chi_I$. They correspond to
 the second derivatives of the pressure with respect to $\mu_q$ 
 and $\mu_I$, where $\mu_I$ is the isospin chemical potential. 
 From a phenomenological argument in the sigma model, 
 $\chi_q$ is singular at the critical point, 
 whereas $\chi_I$ shows no singularity there \cite{Hatta}. 

\begin{figure}[tbp]
  \begin{center}
    \begin{tabular}{cc}
    \includegraphics[width=65mm]{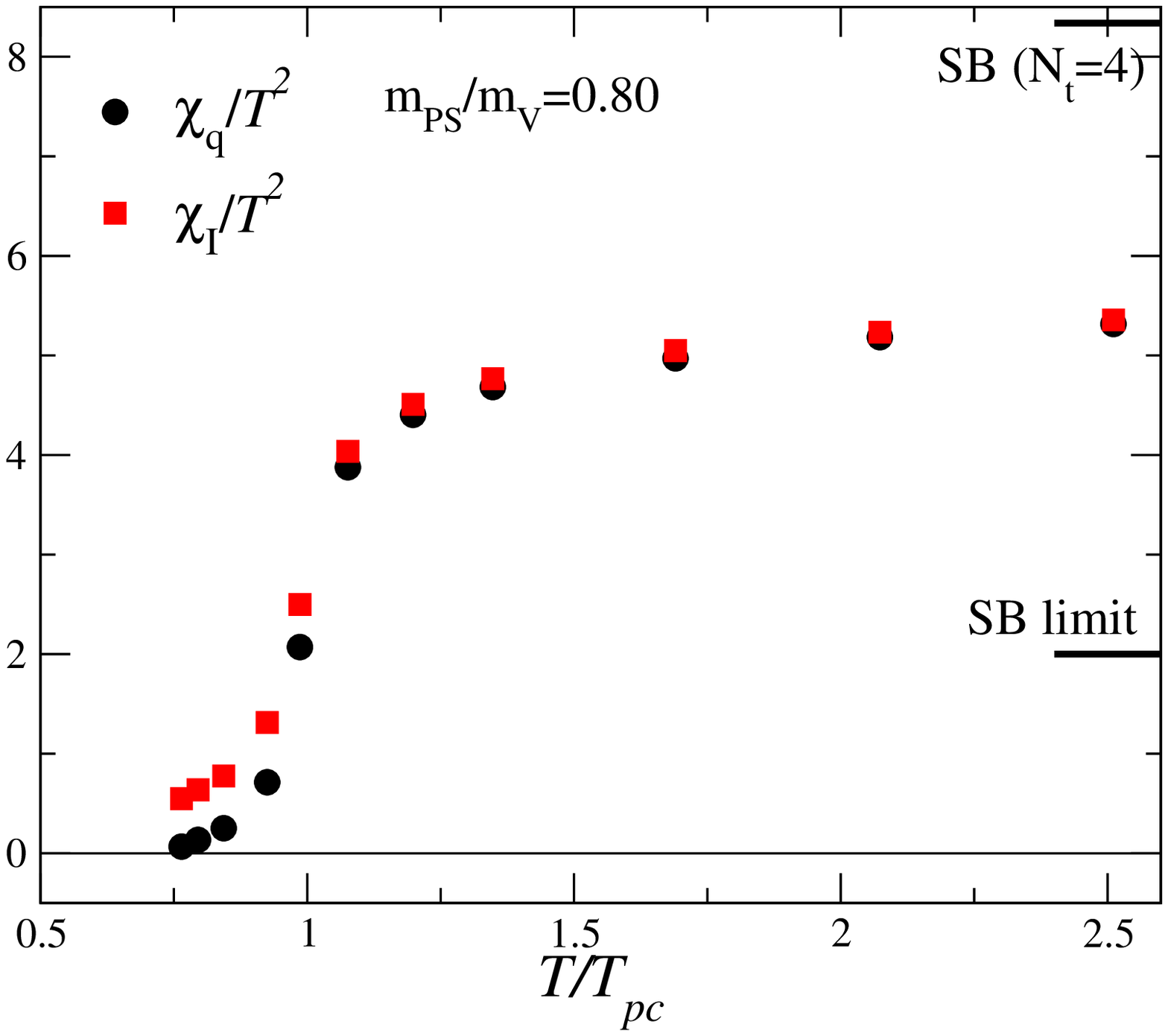} &
    \includegraphics[width=65mm]{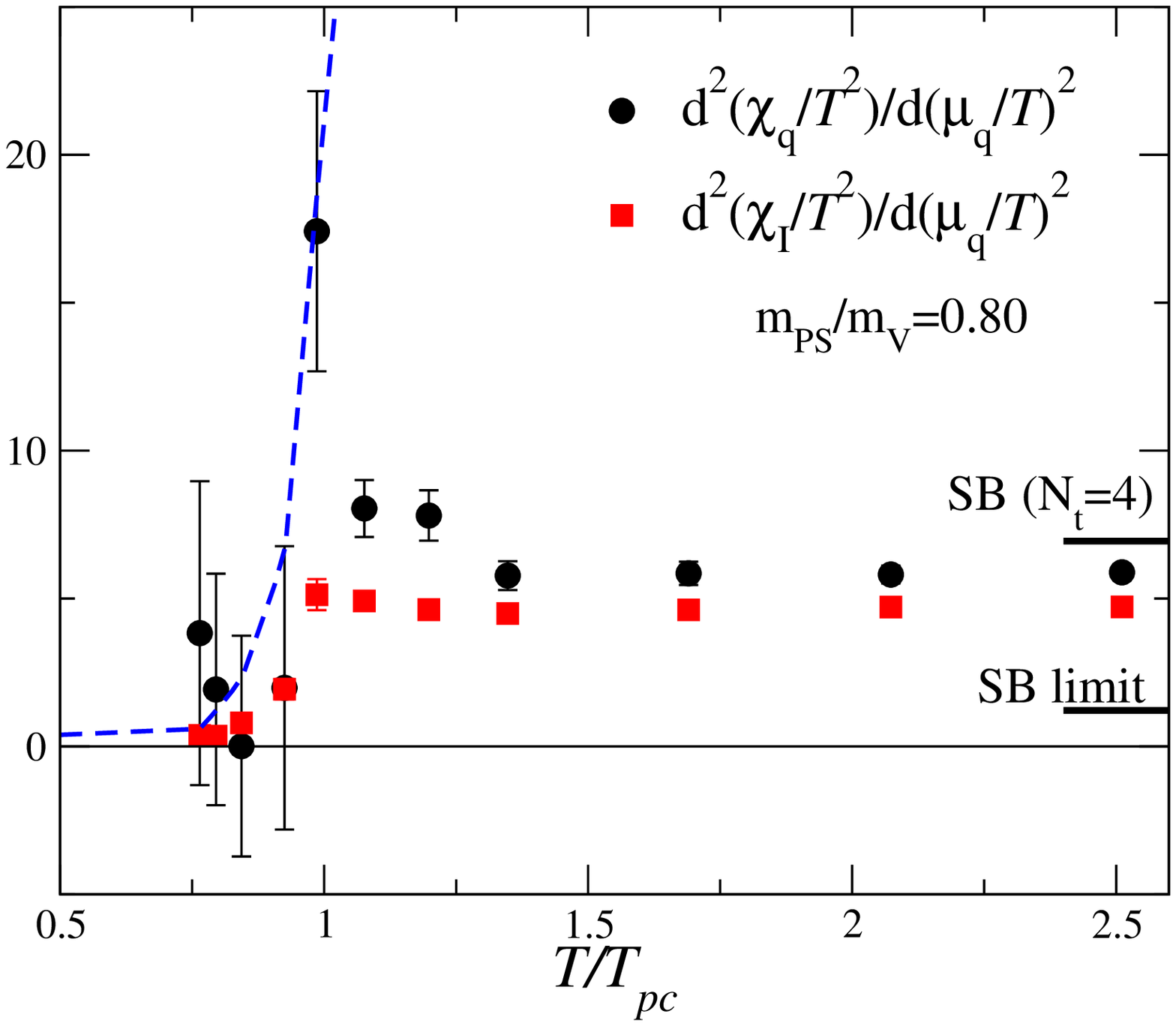}
    \end{tabular}
    \caption{Left: Quark number (circle) and isospin (square)
    susceptibilities at $\mu_q = \mu_I = 0$.
    Right: The second derivatives of these susceptibilities.
    }
 \vspace{-0.5cm}
    \label{fig:qns}
  \end{center}
\end{figure}

In this section, 
we calculate the $\chi_q$ and $\chi_I$ and their second derivatives  
with respect to $\mu_q$ and $\mu_I$ 
at $\mu_q = \mu_I =0$. (Note that the odd derivatives are zero at $\mu_q=0$.) 
The details of the calculations are reported in Ref.~\cite{ejiri}.
The left panel of Fig.~\ref{fig:qns} shows 
 $\chi_q/T^2$ (circle) and $\chi_I/T^2$ (square) 
 at $m_{\rm PS}/m_{\rm V}=0.8$ and $\mu_q = \mu_I =0$ as functions of $T/T_{pc}$.
We find that $\chi_q/T^2$ and $\chi_I/T^2$ 
increase sharply at $T_{pc}$, in accordance with the expectation 
that the fluctuations in the QGP phase are much larger 
than those in the hadron phase.
Their second derivatives $\partial^2 (\chi_q/T^2)/ \partial (\mu_q/T)^2$ 
and $\partial^2 (\chi_I/T^2)/ \partial (\mu_q/T)^2$ are shown 
in Fig.~\ref{fig:qns} (right). 
We find that basic features are quite similar to those found previously with the  p4-improved staggered fermions \cite{ks}. 
$\partial^2 (\chi_I/T^2)/ \partial (\mu_q/T)^2$ remains small around $T_{pc}$,
suggesting that there are no singularities in $\chi_I$ at non-zero density. 
On the other hand, we expect a large enhancement in the quark number 
fluctuations near $T_{pc}$ as approaching the critical point in the 
$(T, \mu_q)$ plane. 
The dashed line in Fig.~\ref{fig:qns} (right) 
is a prediction from the hadron resonance gas model,
  $\partial^2 \chi_q/ \partial \mu_q^2 \approx 9 \chi_q/T^2$.
Although  current statistical errors 
in Fig.~\ref{fig:qns} (right) are still large, 
we find that  
$\partial^2(\chi_q/T^2) / \partial (\mu_q/T)^2$ near $T_{pc}$ is 
much larger than that at high temperature.
At the right end of the figure, values 
of free quark-gluon gas (Stefan-Boltzmann gas) 
 for $N_t=4$ and for $N_t = \infty$ limit are shown.
 Since the lattice discretization error in the equation of state is 
 known to be large at $N_t=4$ with our quark action,
 we need to extend our study to larger $N_t$ for the continuum extrapolation. 

%%%%%%%%%%%%%%%%%%%%%%%%%%%%%%%%%%%%%%%%%%%%%%%%%%%%%%%%%%%%%%%%%%%%%
\section{Summary}

We presented current status of thermodynamics 
of two-flavor QCD with the renormalization group improved gauge action 
and the clover improved Wilson quark action. 
Simulations were performed on a $16^3 \times 4$ lattice
and along lines of constant $m_{\rm PS} / m_{\rm V} = 0.65$ and 0.80.

The properties of the heavy-quark free energies 
at finite $\mu_q$ were studied
in the Taylor expansion method up to 2nd order of $\mu_q$.
We find that
there is a characteristic difference between $Q\bar{Q}$
  and $QQ$ free energies
  due to the first order coefficient of the Taylor expansion.
 It suggests that the inter-quark interaction between $Q$ and $\bar{Q}$
 ($Q$ and $Q$) become week (strong) in the leading-order
 of $\mu_q$.
 We also extract
 the expansion coefficients of the Debye screening mass for each
 color channel up to 2nd order of $\mu_q$.
The Debye mass shows no significant channel dependence
 at $T \simge 2 T_{pc}$, whereas, we find disagreement with 
 leading-order predictions of the thermal perturbation theory.
Since it is known that the next-to-leading-order 
  of the thermal perturbation with respect to $T$
  well reproduces the Debye mass on a lattice 
at $\mu_q = 0$ \cite{whot}, 
the calculations of the higher order perturbation
with respect to $\mu_q$ 
will help us to understand the results on the lattice.
 
The fluctuations of quark number and isospin densities were also discussed.
Although the statistical errors are still large, we find that 
$\chi_q$ seems to increase rapidly near $T_{pc}$ as $\mu_q$ increases, 
whereas the increase of $\chi_I$ is not large near $T_{pc}$. 
These behaviors qualitatively agree with the previous 
results obtained with the p4-improved staggered fermions.

\paragraph{Acknowledgements:}
This work is in part supported by Grants-in-Aid of the Japanese 
MEXT
(Nos.~13135204, 15540251, 17340066, 18540253).
YM is supported by JSPS, and
SE is supported by the U.S. Department of Energy (DE-AC02-98CH1-886).
This work is in part supported also by ACCC, Univ. of Tsukuba, 
and the Large Scale Simulation Program No.06-19 (FY2006) of
 KEK.


\begin{thebibliography}{99}

\bibitem{cp}
  A.~Ali Khan {\it et al.}  (CP-PACS Collaboration),
  \emph{Phys.~Rev.~D}{\bf 63} (2001) 034502;
 {\bf 64} (2001) 074510.

\bibitem{whot}
  Y.~Maezawa {\it et al.} (WHOT-QCD Collaboration), 
  \emph{Phys.~Rev.~D}{\bf 75}, (2007) 074501.

\bibitem{ejiri}
  S.~Ejiri {\it et al.}
in proceedings of \emph{Lattice 2006}.
\pos{PoS(LAT2006)132}.


\bibitem{ks1}
M. D\"{o}ring, S. Ejiri, O. Kaczmarek, F. Karsch and E. Laermann,
  \emph{Eur.~Phys.~J.~C}{\bf 46}, 179 (2006).

\bibitem{Gockeler:2005rv}
  M.~Gockeler {\it et al.},
  \emph{Phys.~Rev.~D}{\bf 73} (2006) 014513.

\bibitem{Hatta}
  Y.~Hatta and M.~A.~Stephanov,
  \emph{Phys.~Rev.~Lett.}{\bf 91} (2003) 102003.


\bibitem{ks}
  C.R. Allton {\it et al.},
   \emph{Phys.~Rev.~D}{\bf 68} (2003) 014507;
  {\bf 71} (2005) 054508.
  




\end{thebibliography}
\end{document}